\documentclass[journal]{IEEEtran}

\usepackage[utf8]{inputenc}
\usepackage[T1]{fontenc}
\usepackage{graphicx}
\usepackage{grffile}
\usepackage{longtable}
\usepackage{wrapfig}
\usepackage{rotating}
\usepackage[normalem]{ulem}
\usepackage{amsmath}
\usepackage{textcomp}
\usepackage{amssymb}
\usepackage{capt-of}
\usepackage{hyperref}
\usepackage{color}
\usepackage{listings}
\usepackage[utf8]{inputenc}
\usepackage[T1]{fontenc}
\usepackage{graphicx}
\usepackage{amsmath, braket}
\usepackage{cite}

\begin{document}

\title{Robust Electric-field Input Circuits for Clocked Molecular Quantum-dot Cellular Automata}

\author{
\IEEEauthorblockN{
   Peizhong Cong\IEEEauthorrefmark{1} and
   Enrique P. Blair\IEEEauthorrefmark{1}}
   
\thanks{This work was funded by the Office of Naval Research under grant
N00014-20-1-2420.}
\thanks{Peizhong Cong and Enrique P. Blair are with the Electrical and Computer Engineering Department, Baylor University, Waco, TX 76798 USA (e-mail: \href{mailto:joe_cong1@baylor.edu}{joe\_cong1@baylor.edu}; \href{mailto:Enrique_Blair@baylor.edu}{Enrique\_Blair@baylor.edu})}

\IEEEauthorblockA{\IEEEauthorrefmark{1}
Electrical and Computer Engineering Department, Baylor University, Waco, Texas, United States of America}
}

\IEEEtitleabstractindextext{

\begin{abstract}
Quantum-dot cellular automata (QCA) is a paradigm for low-power,
general-purpose, classical computing designed to overcome the
challenges facing CMOS in the extreme limits of scaling. A molecular
implementation of QCA offers nanometer-scale devices with device 
densities and operating speeds which may surpass CMOS device densities
and speeds by several orders of magnitude, all at room
temperature. Here, a proposal for electric field bit write-in to
molecular QCA circuits is extended to QCA circuits clocked
using an applied electric field, \(\vec{E}\). Input electrodes, which may be much larger than the cells 
themselves, immerse an input circuit in an input field \(E_y \hat{y}\), in
addition to the applied clocking field \(E_z \hat{z}\). The input
field selects the input bit on a field-sensitive portion of the
circuit. Another portion of the circuit with reduced \(E_y\)-sensitivity
functions as a shift register, transmitting the input bit to
downstream QCA logic for processing. It is shown that a simple rotation
of the molecules comprising the shift register makes them immune to
unwanted effects from the input field or fringing fields in the
direction of the input field. Furthermore, the circuits also tolerate
a significant unwanted field component \(E_x \hat{x}\) in the third
direction, which is neither the clocking nor input direction. The
write-in of classical bits to molecular QCA circuits is a road-block
that must be cleared in order to realize energy-efficient molecular
computation using QCA. The results presented here show that
interconnecting shift registers may be designed to function in the
presence of significant unwanted fringing fields from large input
electrodes. The techniques devloped here may also enable
molecular QCA logic to tolerate these same unwanted fringing fields.\end{abstract}

\begin{IEEEkeywords}
Molecular Quantum-dot Cellular Automata, synchronous input circuit, fringing fields, electric-field input, write-in, clocked QCA
\end{IEEEkeywords}}

\maketitle

\IEEEdisplaynontitleabstractindextext

\IEEEpeerreviewmaketitle

\section{Introduction}
\label{sec:org422d752}
Quantum-dot cellular automatata (QCA) is a low-power, energy-efficient
paradigm for general-purpose computing in the beyond-CMOS era
\cite{LentTougawPorodBernstein:1993,Snider1998}. A molecular 
implementation of QCA promises nanometer-scale devices with device
densities approaching \(10^{14}\; \text{cm}^{-2}\) and with
\(\sim\)THz switching speeds at room temperatures
\cite{LentScience2000,2003-JACS-mQCA,molecularQCAelectronTransfer}.

One challenge in molecular QCA is the write-in of classical bits on
nano-scale molecules. While some techniques have been proposed for
bit-write-in to molecular QCA circuits
\cite{WalusQCAInput,2012-nanoelectrodes}, we have proposed a technique
that requires neither special fixed-state QCA molecules nor electrodes
with single-molecule specificity \cite{QCAeFieldWriteIn-IEEE}. Quantum
models of asynchronous QCA circuits were used to demonstrate that
charged electrodes much larger than the molecules could be used to
apply an input electric field to the molecular circuitry and
write bits onto several molecules. Furthermore, it was shown that
molecular QCA interconnections may tolerate significant unwanted
fringing fields from the input electrodes.

In this paper, the previous model for unclocked molecular QCA input
circuits is extended to circuits comprised of clocked, three-dot
molecules.  We calculate the ground state response for such input circuits in both an idealized limit
(without unwanted field fringing), and in an extreme case
of fringing fields that are as strong as the intended 
input field. Clocked input circuits are shown to function
properly\textemdash that is, the ground state encodes the correct calculational result\textemdash even under significant unwanted fringing fields. We also
demonstrate that parts of the circuits may be made even more 
robust to fringing fields by a simple rotation of the cells by 90
degrees. The rotated cells exhibit immunity to fringing fields in the
direction perpendicular to their longitudinal axis, as well as significant
robustness against fringing fields parallel to their longitudinal
axis. This may provide a simple means for insulating QCA logic and
interconnects from unwanted fringing fields from input electrodes.

This paper provides an overview of QCA in Section
\ref{sec:org421cc4f}, with a focus on molecular QCA and 
clocking. Section \ref{sec:org51a1bb7} introduces a three-state 
Hamiltonian for an individual molecular QCA device under the influence of
clocking and input fields. While previous work treated input circuits using an intercellular Hartree approximation (ICHA)
\cite{mQCA_inputs_ICHA}, here, we form the full \(3^N\)-dimensional Hamiltonian for an \(N\)-device circuit, and we find its ground state. A clocked molecular input circuit is presented. Then, in Section
\ref{sec:org5b05c69}, the ground state of the clocked input circuit is
shown to properly encode the desired logical response, even under significant external electric fields. Additionally, a more robust design for the input circuit is
proposed, and its response is calculated and reported. We also briefly discuss in Section \ref{sec:Discussion} other challenges to the realization of molecular QCA computation, as well as promising solutions to these challenges.

% ============================================================
% ============================================================
% ============================================================
% SECTION: Overview of QCA
% ============================================================
% ============================================================
% ============================================================
\section{Overview of QCA}
\label{sec:org421cc4f}
\begin{figure}[htbp]
\centering
\includegraphics[width=.9\linewidth]{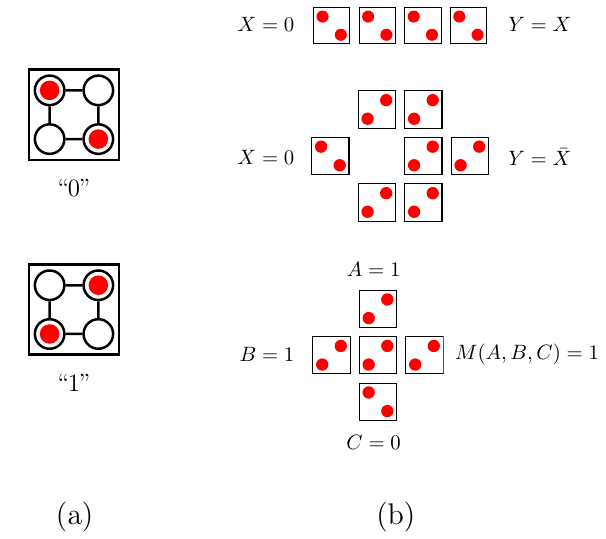}
\caption{(a) Two charge configurations of mobile electrons (red discs) on a four-dot QCA cell encode a bit. The black circles schematically represent quantum dots, and black lines connecting dots indicate a tunneling path. (b) A universal set of QCA logic circuits may be formed by coupling cells locally through the Coulomb field. The location of a cell is marked by a black square. Only the location of charge is shown (dots are not drawn). Shown here are the binary wire (top), the inverter (middle), and the majority gate (bottom). \label{four-dot-cells-logic}}
\end{figure}

In QCA, a classical bit is encoded
in the configuration of a few mobile charges on a system of quantum
dots called a \emph{cell}. A four-dot cell with two electrons is shown schematically in
Figure \ref{four-dot-cells-logic}(a). The isolated cell has two
degenerate, localized electronic states, which are assigned binary
labels ``0'' and ``1.'' Device switching takes place via quantum
tunneling of mobile charge between dots.

Neighboring cells couple locally via the Coulomb field so that
logic circuits may be formed by arranging cells on a substrate.
A logically complete set of circuits is shown in Figure
\ref{four-dot-cells-logic}(b) 
\cite{LentTougawPorodBernstein:1993,TougawLentLogDev1994}. These include:
a binary wire, in which input bit \(X\) is copied from one cell to the
next, producing output \(Y=X\); an inverter, 
which uses diagonal coupling to achieve a bit flip to obtain
output \(Y=\bar{X}\); and the majority gate, which may be used to implement
a programmable, two-input, AND/OR gate. Here, the central 
device cell takes the state in the majority of three inputs, \(A\),
\(B\), and \(C\) and then copies that bit, \(M(A,B,C)\), to the
output.  Much more complex circuits have been designed from these building
blocks, up to entire processors \cite{Simple12Conf}.

QCA have been implemented in various ways. The earliest implementation
used metal islands patterned on an oxide for quantum dots
\cite{Snider1998}. Later implementations used semiconductor dots
\cite{Smith2003,SiQCA2003} or dangling bonds on a Si surface as atomic-scale dots
\cite{WolkowQCA_Silicon}.

Various QCA devices have been fabricated. These include cells \cite{QCACell1999,1999ClockingToth,SiQCA2003,Smith2003,LeadlessQCA,WolkowQCA_Silicon},
latches \cite{QCALatch},
binary wires \cite{Orlov1999}, majority gates \cite{ScienceQCALogic1999}, and shift registers \cite{PowerGainShiftReg2001}.

\begin{figure}[htbp]
\centering
\includegraphics[width=.9\linewidth]{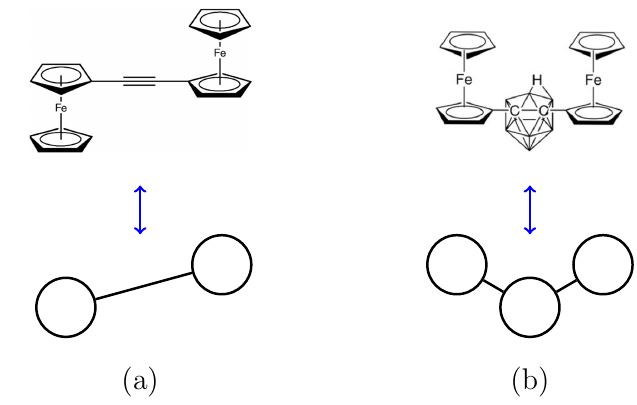}
\caption{Examples of QCA molecules. (a) A diferrocenyl acetylene (DFA) molecule [top] has two Fe centers, which provide one redox center each and function as a quantum dot. The DFA molecule, then, is a double-quantum-dot system [bottom, shown schematically], which may provide half of a four-dot cell. (b) A zwitter-ionic nidocarborane molecule [top] provides a three-dot system [bottom], which functions as a clocked QCA cell. This may provide half of a six-dot QCA cell. \label{fig-qca-molecules}}
\end{figure}

This paper 
focuses on molecular QCA \cite{LentScience2000,mQCA_00}. Here, an
individual mixed-valence molecule functions as a cell, with redox
centers on the molecule providing dots. One example of a QCA molecule
is the diferrocenyl acetylene (DFA)
\cite{DFA,molecularQCAelectronTransfer}, as shown in Figure 
\ref{fig-qca-molecules}(a). Figure \ref{fig-qca-molecules}(b) shows a
zwitterionic nido-carborane molecule \cite{ZwitterionicNidoCarborane},
which provides a three-dot QCA cell. Like the atomic-scale dots \cite{WolkowQCA_Silicon},
molecular QCA are helpful implementations because they allow bit
energies that are robust at room temperature, eliminating the need for
cryogenic cooling.

\begin{figure}[htbp]
\centering
\includegraphics[width=.9\linewidth]{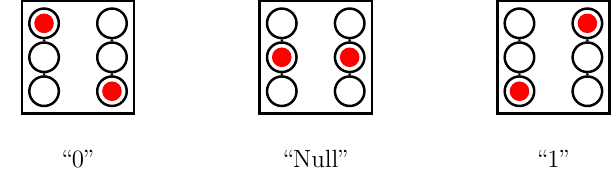}
\caption{A six-dot cell with an additional ``Null'' state enables clocking. \label{six_dot_states}}
\end{figure}

An important development in QCA is clocking, which enables the
latching of bits onto cells \cite{clocked_qca_latch}, synchronous calculations, adiabatic circuit operation \cite{1999ClockingToth,BennettClocking,ExponentiallyAdiabaticSwitch2018}, and provides
power gain for the restoration of weakened signals
\cite{Timler2002}. This may be achieved by adding a third pair of dots,
as in Figure \ref{six_dot_states}. The new dots provide a ``Null'' state
with no information content. A potential may be applied to the central
dots to clock the six-dot cell to the ``Null'' state; or, the polarity of that
potential may be reversed to repel the mobile charge from the central
null dots, clocking the six-dot cell to an active state, ``0'' or ``1.'' To
switch a six-dot cell between active states requires an intermediate
transition to the ``Null'' state. Thus, clocking a six-dot cell to an active
state suppresses tunneling between active states and latches the bit
on that six-dot cell. Logic may be formed from clocked six-dot cells as
previously discussed.

In molecular QCA, a pair of three-dot molecules like the one shown in Figure \ref{fig-qca-molecules}(b) could be juxtaposed to function as a molecular clocked six-dot cell, as in Figure \ref{six_dot_states}. The component of an applied electric field perpendicular to the
device plane (\(\vec{E}_z \hat{z}\)) may be used to clock the molecular devices and circuits
\cite{Hennessy2001,BlairLentArchitecture,2018_clock_topologies}.

In the remainder of this paper, we focus on three-dot molecules only. Three-dot molecules are useful because they can be arranged in circuits which interact with electric field like dipoles, providing a circuit with high sensitivity to the applied \(\vec{E} (\vec{r})\); or they may be arranged to interact only weakly with the applied \(\vec{E} (\vec{r})\) as quadrupoles.  Additionally, we use the terms ``cell'' and ``molecule'' interchangeably.

% ============================================================
% ============================================================
% ============================================================
% SECTION: Overview of QCA
% ============================================================
% ============================================================
% ============================================================
\section{Model}
\label{sec:org51a1bb7}

The objective here is to explore the design and function of clocked QCA input circuits in the presence of unwanted electric fields. To do this, we develop a model for a circuit of clocked, three-dot molecular QCA immersed in an external electric field, \(\vec{E}\). The time-independent Schr\"odinger equation is solved for the ground state of the circuit. The circuit will have succeeded when its ground state encodes the intended calculational result, and it will have failed when the ground state diverges from the correct output. First, the Hamiltonian, \(\hat{H}_k\), of a single cell is introduced and validated in subsection \ref{sec:org03a42df}. Then, in subsection \ref{sec:orge7e4c81}, the non-interacting Hamiltonian, \(\hat{H}_0\), for an \(M\)-cell circuit is formed by summing all single-cell Hamiltonian operators \(\{ \hat{H}_k \}\). The circuit Hamiltonian, \(\hat{H}\) is completed by adding to \(\hat{H}_0\) an interaction term, \(\hat{H}_{\text{int}}\). \(\hat{H}_{\text{int}}\) is calculated directly from electrostatics. The clocked input circuit is then introduced in subsection \ref{sec:org8a8a904}. We limit the circuit size to \(M \leq 8\), since larger circuits become intractable due to the exponential growth of the dimension \(d\) of the circuit Hilbert space with \(M\): \(d = 3^M\). Given the relatively small size of the molecular circuits treated here, we assume that the clocking electric field \(\vec{E}_z \hat{z}\) is uniform over the entire circuit.

% ============================================================
% ============================================================
% SUBSECTION: Molecular Device Description
% ============================================================
% ============================================================
\subsection{Molecular Device Description}
\label{sec:org03a42df}
\begin{figure}[htbp]
\centering
\includegraphics[width=.9\linewidth]{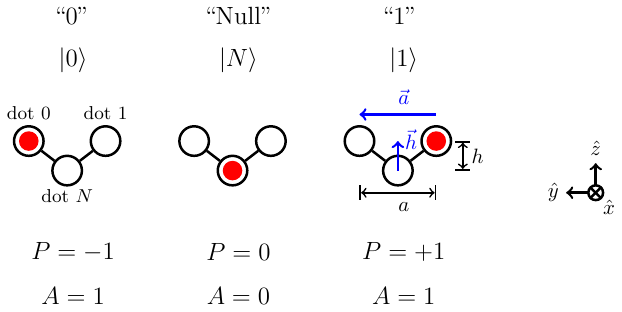}
\caption{Localized configurations of a mobile electron (red disc) on one of three molecular quantum dots provides three distinct device states. \label{fig-three-dot-states}}
\end{figure}

Figure \ref{fig-three-dot-states} provides a schematic depiction of a
three-dot cell. The three molecular dots (black circles) are labeled
0, 1, and \(N\). The state \(\ket{x}\) is the localized state of one
mobile electron (red disc) on dot \(x \in {0, 1, N}\). These are assigned device
values ``0,'' ``1,'' and ``Null,'' respectively. States ``0'' and
``1'' represent a bit and are designated as ``active states'' of the
molecular cell. Not depicted here is a fixed neutralizing charge,
assumed to be located at the null dot. The neutralizing charge
\(+q_e\) gives the molecule net charge neutrality, modeling a
zwitterionic QCA molecule. For simplicity, the cell is treated as a
system of two point charges: the mobile charge, \(-q_e\), and the fixed
charge, \(+q_e\), where \(q_e\) is the elementary charge. Black lines
connecting dots represent tunneling paths for the mobile 
charge. As discussed above, direct tunneling between active states is
suppressed, enabling the latching of a bit onto a cell. The active
dots are separated by distance \(a\). Here, we will 
consider molecules adsorbed onto a surface (the \(z=0\) plane) so that
dot \(N\) is on the surface and active dots are elevated to the plane
\(z = +h\).

Consider an isolated cell at position \(\vec{r}\) immersed in an
electrostatic field \(\vec{E} (\vec{r})\), which is approximately
constant over the volume of the cell. The Hamiltonian of the cell may
be written as
\begin{align}
\hat{H} & = -\gamma \left( \hat{P}_{0,N} + \hat{P}_{N,0} +
\hat{P}_{1,N} + \hat{P}_{N,1} \right) \nonumber \\
& \qquad + \frac{\Delta}{2} \hat{Z} + \left( V_c - E_a \right) \hat{P}_N \, , \label{eq-H-single-cell}
\end{align}
where \(\hat{P}_{\alpha,\beta} \equiv \ket{\alpha} \bra{\beta}\) is a
transition operator from \(\ket{\beta}\) to \(\ket{\alpha}\), and \(\hat{P}_{\alpha} \equiv \hat{P}_{\alpha,
\alpha}\) is a projection operator onto \(\ket{\alpha}\). The operator \(\hat{Z} \equiv \hat{P}_1 -\hat{P}_0\) is analogous to the Pauli
operator \(\hat{\sigma}_z\) and employs the same sign convention for
\(\hat{\sigma}_z\) as in Ref. \cite{QNetworksMahlerWeberuss}. The first
term in Equation (\ref{eq-H-single-cell}) describes electronic tunneling
between the active states and the ``Null'' state, and is characterized 
by \(\gamma\), the hopping energy between either of the active states
and the ``Null'' state. It is assumed that the molecule is symmetric so
that \(\gamma\) describes both the \(\ket{0}\rightarrow \ket{N}\)
transition and the \(\ket{1}\rightarrow \ket{N}\) transition. The
second term in Equation (\ref{eq-H-single-cell}) is characterized by
\(\Delta\), the detuning between states ``0'' and ``1'':
\begin{equation}
\Delta = \Braket{1 | \hat{H} | 1} - \Braket{0 | \hat{H} | 0} \; .
\end{equation}
The detuning may be separated into two components:
\begin{equation}
\Delta = \Delta_{\text{Neigh}} + \Delta_E \; ,
\end{equation}
where \(\Delta_{\text{Neigh}}\) is the bias due to the charge
distribution of neighboring QCA cells, and \(\Delta_E\) is a bias
driven by the applied field, \(\vec{E}\). The field-driven bias is
given by
\begin{equation}
\Delta_E = - q_e \cdot \vec{E} (\vec{r}) \cdot \vec{a} \;,
\end{equation}
where \(\vec{a}\) is the vector of length \(a\) pointing from dot 1 to
dot 0 (see Figure \ref{fig-three-dot-states}). The third term of Equation 
(\ref{eq-H-single-cell}) is proportional to
\(\hat{P}_N\) and describes the occupation energy of the ``Null'' state
\(\ket{N}\) relative to the unperturbed active states. Here, the clock
biases the ``Null'' state relative to the active states by the potential energy
\[ V_c = -q_e \vec{E} \cdot \vec{h},\]
and the energy \(E_a\) describes the affinity the mobile charge has
for the fixed neutralizing charge on dot \(N\). An \(E_a > 0\) biases
the system toward the ``Null'' state. \(E_a\) is a property of a
particular QCA molecule, which may be estimated using quantum
chemistry techniques or measured experimentally.

A cell's charge state may be characterized by two real numbers: its
polarization, \(P\), and its activation, \(A\):
\begin{align}
P & = \braket{\hat{Z}}, \qquad \mbox{and} \label{eqn-Pol} \\
A & = 1 - \braket{\hat{P}_N}. \label{eqn-Act}
\end{align}
The sign of \(P\) encodes a classical bit, and \(A\) may be understood
as the probability that the molecule will be measured in an active
state.

\begin{figure}[htbp]
\centering
\includegraphics[width=.9\linewidth]{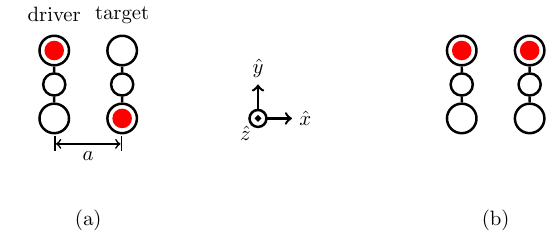}
\caption{(a) The relaxed configuration of the target cell is state ``1'' in the presence of a diver in state ``0.'' (b) The kinked state of the target cell is state ``0'' in the presence of the driver. \label{three-dot-kink}}
\end{figure}

We introduce the kink energy, \(E_k\), which is interpreted as the
cost of a bit flip or the strength of a bit \cite{1994PhysCompLentTougawPorod}. Consider a driver cell of characteristic length \(a\)
prepared in state \(\ket{0}\).  The kink energy for a target cell a
distance \(a\) away from the driver cell is the difference in energy
between the target cell's frustrated state and its relaxed state. Figure
\ref{three-dot-kink}(a) shows the relaxed state, which is anti-aligned
with the driver. In its frustrated state, the target cell aligns with
the driver cell, as in Figure \ref{three-dot-kink}(b). The kink energy
may be calculated directly using electrostatics: 
\begin{equation}
E_k = \frac{q_e^2}{4\pi \epsilon_0 a} \left( 1 - \frac{1}{\sqrt{2}}\right),
\label{eq:kink_energy}
\end{equation}
where \(\epsilon_0\) denotes the permittivity of free space.

In this paper, all calculations assume parameters \(a = 1~\text{nm}\),
\(h=a/2\), \(\gamma = 50~\text{meV}\), and \(E_a =
1~\text{eV}\). These parameters are chosen to represent a molecule
like that of Figure \ref{fig-qca-molecules}(b). It is also
assumed that the mobile charge is one electron, and that the fixed
neutralizing charge is positive. For \(a = 1~\text{nm}\), \(E_k =
422~\mbox{meV}\).

\begin{figure}[htbp]
\centering
\includegraphics[width=.9\linewidth]{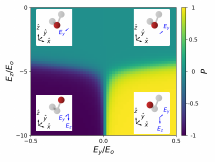}
\caption{An isolated cell responds to the applied field, \(\vec{E}\). The \(z\)-component of \(\vec{E}\) serves as a clock, and the \(y\)-component of \(\vec{E}\) is termed the ``input field,'' since it may be used to select a bit. \label{fig-field-cell-response}}
\end{figure}

The isolated cell's ground state response \(P\) to various applied fields \(\vec{E}\)
is shown in Figure \ref{fig-field-cell-response}. In this plot, both the
clocking component (\(E_z\)) and the biasing (input) component (\(E_y\)) of
\(\vec{E}\) are scaled to \(E_o\), the field strength required to
produce a kink in two cells, as in Figure \ref{three-dot-kink}(b):
\begin{equation}
E_o = \frac{E_k}{q_e a} = \frac{q_e}{4\pi \epsilon_0 a^2} \left( 1 - \frac{1}{\sqrt{2}} \right) \; .
\end{equation}
In the absence of a clock (\(E_z = 0\)), or under weak clocking (\(E_z
\gg -5 E_o\)), the cell remains in the null state (green region of the
plot) because of the positive electron-hole affinity, \(E_a\). Therefore, some \(E_z < 0\) is required to clock
the cell to an active state. If \(|E_z|\) is large enough, even a weak
input field (\(| E_y | \ll E_o\)) is adequate to select a bit on the
cell (yellow and blue regions of the plot).

\begin{figure}[htbp]
\centering
\includegraphics[width=.9\linewidth]{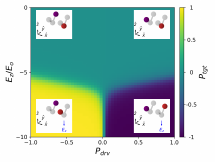}
\caption{A target cell responds to a neighboring driver molecule. Here, only a clocking field \(E_z\) is applied, and the input field is suppressed \(E_y = 0\). In the insets, the mobile charge of the driver cell is colored purple for the sole purpose of identifying the driver cell.  \label{fig-driver-cell-response}}
\end{figure}

A QCA cell also may be driven by a neighboring molecule, as in
Figure \ref{fig-driver-cell-response}. Here, the response is calculated
for a target cell in the presence of a fully-activated
driver cell (\(A_{\text{drv}}=1\)).  % For this calculation, the on-site energies for the target cell's mobile electron in the basis states \(\{\ket{0}, \ket{N}, \ket{1} \}\) have a contribution from to the electrostatic potential established by the driver cell's charge configuration.
The target molecule's response, \(P_{\text{tgt}}\), is plotted
as a function of the driver polarization, \(P_{\text{drv}}\), and
the clocking electric field, \(E_z\). The \(y\)-component of the field
is zero here, since no external input field is applied. The repulsion by the driver cell's mobile charge and the attraction to
the driver cell's neutralizing charge further bias the target cell
toward the null state so that a stronger clock than before is required
to activate the target cell. This effect has been termed ``population
congestion,'' and has been studied in detail \cite{PopCongest2020}.

% ============================================================
% ============================================================
% SUBSECTION: Molecular Circuit Model
% ============================================================
% ============================================================
\subsection{Molecular Circuit Model}
\label{sec:orge7e4c81}
A convenient basis, \(\{\ket{\mathbf{x}}\}\), for an \(M\)-cell circuit may be formed by taking
the direct products of localized single-cell states:
\begin{equation}
\ket{\mathbf{x}} = \ket{x_M x_{M-1} \cdots x_2 x_1 } = \ket{x_M } \ket{x_{M-1}} \cdots \ket{x_2} \ket{x_1 } \;. 
\end{equation}
Here, \(\mathbf{x}\) represents the \(M\)-trit state \(x_M x_{M-1} \cdots x_2
x_1\), and \(x_k \in \{0_k, N_k, 1_k\}\) labels a state of the \(k\)-th cell.

The circuit Hamiltonian \(\hat{H}\) may be written as
\begin{equation}
\hat{H} = \sum_{k=1}^M \hat{H}_k + \hat{H}_{\text{int}} \, ,
\end{equation}
where \(\hat{H}_k\) is the free Hamiltonian of the \(k\)-th cell, and \(\hat{H}_{\text{int}}\) describes the interactions between
neighboring cells. Here, \(\Delta_{\text{Neigh}}\) is not
included in \(\hat{H}_k\), but rather is accounted for in
\(\hat{H}_{\text{int}}\). In the basis \(\{ \ket{\mathbf{x}} \}\), the
interaction \(\hat{H}_{\text{int}}\) is diagonal and may be
written as:
\begin{equation}
\hat{H}_{\text{int}} = \sum_{\mathbf{x}} U_{\mathbf{x}} \hat{P}_{\mathbf{x}} .
\end{equation}
Here, \(U_{\mathbf{x}}\) is the electrostatic energy of interaction
between all cells given circuit state \(\ket{\mathbf{x}}\):
\begin{align}
U_{\mathbf{x}} & = \Braket{ \mathbf{x}| \hat{H}_{\text{int}} | \mathbf{x} } \nonumber \\
& = \frac{1}{4\pi \epsilon_o }\sum_{j=1}^{M-1} \sum_{k>j}^M \sum_{\ell, \ell^{\prime}} \frac{q_{\ell}^{(j)} q_{\ell^{\prime}}^{(k)}}{\left| \mathbf{r}_{\ell}^{(j)} - \mathbf{r}_{\ell^{\prime}}^{(k)} \right| } \; . 
\end{align}
Integers \(j\) and \(k\) label a cell in the circuit, and they are chosen
to avoid double-counting pairwise intercellular interactions or
including unphysical, infinite self-interaction energies. The \(m\)-th
cell is treated as a system of point charges \(\{ q_{\ell}^{(m)}\}\),
with charges at positions \(\{ \mathbf{r}_{\ell}^{(m)} (x_m)\}\). In
the case of the mobile charges, \(\mathbf{r}_{\ell}^{(m)} (x_m)\)
depends on the
state \(x_m \in \{0_m, N_m, 1_m\}\) of cell \(m\). Positive integers \(\ell\)
and \(\ell^{\prime}\) index the point charges consituting cells \(j\)
and \(k\), repsectively. 

% ============================================================
% ============================================================
% SUBSECTION: Clocked Input Circuits
% ============================================================
% ============================================================
\subsection{Clocked Input Circuits}
\label{sec:org8a8a904}

\begin{figure*}
\centering
\includegraphics[width=.9\linewidth]{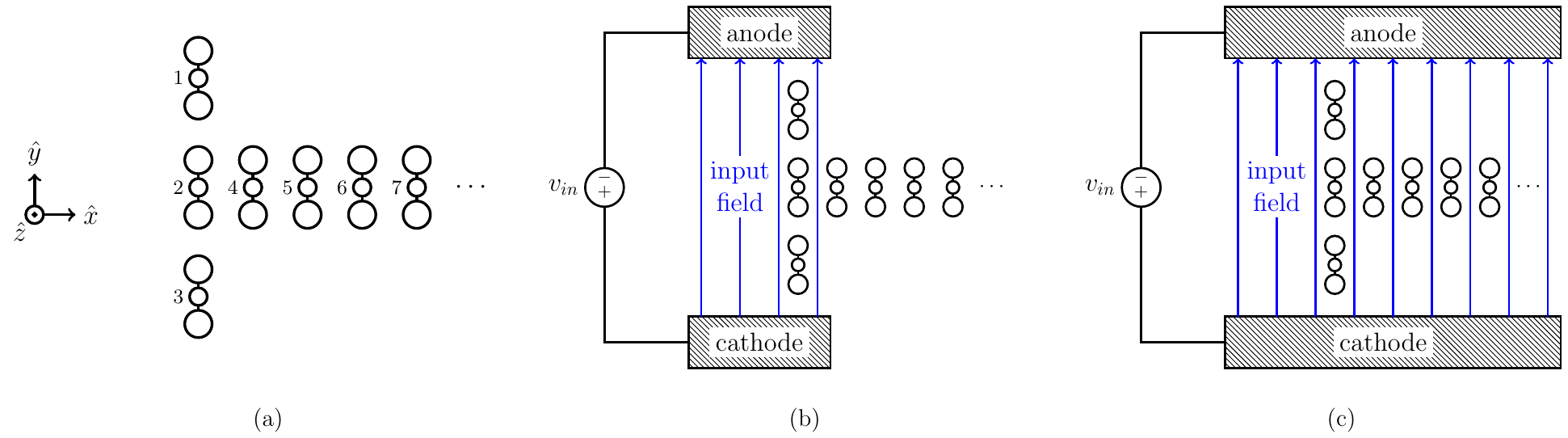}
\caption{A clocked molecular QCA input circuit is studied in two limiting cases. (a) The clocked input circuit is comprised of a field-sensitive component (cells 1-3), and an output shift register (cells 4+). The \(y\)-component, \(E_y\) of an applied electric field, \(\vec{E}\), selects the input bit. The clocked input circuit is studied here in two extreme limits: the idealized nano-electrode limit of subfigure (b), in which the input field is constrained to the input segment; and the large-electrode limit of (c), where the entire input circuit is immersed in a uniform \(E_y\) \label{fig-clocked-input-with-limits}}
\end{figure*}

A clocked molecular QCA input circuit is shown in Figure
\ref{fig-clocked-input-with-limits}(a). The molecular circuit consists
of an input segment (cells 1-3) and a binary wire (cells
4,5,\ldots).

The input segment also is referred to as a
longitudinal array, since the cells are aligned along their
longitudinal axes, the axis which points from one active dot to
another. The minimal field-sensitive segment is two cells long. A longitudinal section of two or more cells has high sensitivity to the field component along its longitudinal axis, since each member cell interacts with the field as a dipole, aligning with both the field and one another. The input segment could be longer than three cells, but we limit its length to three cells in order to limit the dimension of the circuit Hamiltonian, as well as to provide a circuit with symmetry about the \(\hat{x}\) axis. This keeps the calculation manageable, and it allows the circuit to respond in a symmetric way to the input field, \(E_y \hat{y}\).

The binary wire is used to shift the bit selected on the
input segment downstream to QCA logic for processing. The binary wire exhibits some insensitivity to the \(\hat{y}\)-component of the field, since adjacent cells tend to antialign, and pairs of cells interact with the field more weakly as quadrupoles.  We refer to the binary wire as a shift register henceforth, because its constituent
cells are able to latch bits under the clock.

There is no theoretical limit to the length of the shift register. An \textit{un}clocked binary wire, on the other hand, is limited in length, since the entropy of a bit error (a kinked state) increases with the length of the wire, lowering the free energy of such errors \cite{1994PhysCompLentTougawPorod}. For an adequately long wire, the free energy of a kink will fall below that of an error-free wire, giving rise to an error despite bit energies towering high above the thermal noise floor.  A shift register, however, is actively driven by the clock: it is not in thermal equilibrium, and the clock provides power gain to restore weakened signals. Thus, a shift register can be longer than an unclocked wire \cite{2008_DefectTolerance}.
% strong bit energies, numerous possible kinked configurations contribute to a large free energy of kinked states . Once this free energy is grows large enough by virtue of wire length, it makes an error likely, despite a robust kink energy \cite{1994PhysCompLentTougawPorod}. A clocked shift register, however, is actively driven by the clock: it is not in thermal equilibrium, and the clock provides power gain to restore weakened signals. Thus, a clocked shift register can be longer than the unclocked wire \cite{2008_DefectTolerance}.

The clocked input circuit is a synchronous version of the ballistic
input circuit developed previously \cite{QCAeFieldWriteIn-IEEE}. To select the input bit, 
the input circuit is immersed in an applied electric field parallel
with the input section. The field may be established using input
electrodes. This circuit is similar to an input from a proposals in
which electrodes establish an electric field to write a bit onto a
single molecule \cite{2012-nanoelectrodes}, but is designed to support
input electrodes lacking single-molecule specificity. Previously, the
clocked circuit of Figure \ref{fig-clocked-input-with-limits}(a) was
studied \cite{mQCA_inputs_ICHA} using an intercellular Hartree
approximation (ICHA) \cite{TougawLent1996}, which discards intercellular
correlations. This model includes all intercellular correlations and
treats the circuit exactly in the limit of three-state quantum
devices.

In this paper, two limiting cases of the field are studied. First is an
idealized limit, which we call the \emph{nano-electrode limit}. Here, the
input field may be constrained to only the input segment, as
illustrated Figure \ref{fig-clocked-input-with-limits}(b). We also
consider the \emph{large-electrode limit}, in which the entire circuit is
immersed in the input field, as in Figure
\ref{fig-clocked-input-with-limits}(c). This represents the worst case,
in which the \(y\)-component of the unwanted fringing \(\vec{E}\) is
as strong as the input \(E_y\) intentionally applied to the input
portion of the circuit. A more realistic situation is intermediate to
these two limits: input electrodes establish an input
\(|E_{y}^{\left(\text{in}\right)}|\) across the input segment, and
field fringing applies some
\(|E_{y}^{\left(\text{fringe}\right)}|<|E_{y}^{\left(\text{in}\right)}|\) 
to the downstream QCA circuitry. If the circuits function properly in the
large-electrode limit, then it is reasonable to expect that the circuits 
will function properly in the more realistic, intermediate regime.

In this paper, circuits are limited to \(M \leq 8\) cells, since presently, calculations are unwieldy for \(M>8\). Given the small size of these circuits, we apply in all calculations a uniform clock \(E_z \hat{z}\) to all cells, which approximates a clock having a negligible gradient over the circuit modeled.

\section{Results}
\label{sec:org5b05c69}
\subsection{Clocked Input Circuit Response}
\label{sec:org6b61caa}
\begin{figure}[htbp]
\centering
\includegraphics[width=.9\linewidth]{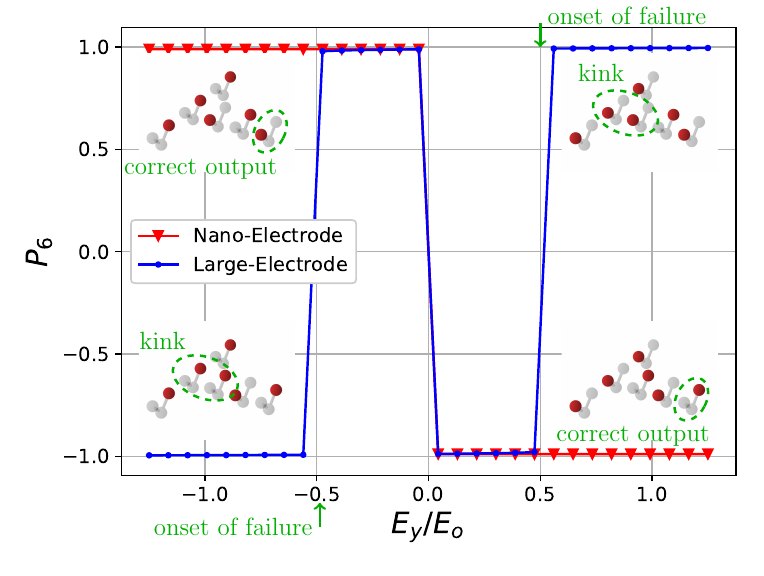}
\caption{An input circuit responds to the clock and an input field. The two limiting cases for the field are shown here: the nano-electrode limit (ideal), and the large-electrode (worst case for fringing fields). In the large-electrode limit, strong unwanted fringing fields \(|E_y| > E_o/2\) disrupt circuit operation. \label{fig-basic-clocked-input-ckt}}
\end{figure}

The ground state response of a clocked input circuit is shown in Figure
\ref{fig-basic-clocked-input-ckt}, and both the idealized and worst-case
of fringing fields are considered. In the ideal case, the sign of the
field selects the output bit of the circuit, which is taken here to be
encoded on cell 6. The same
holds true in the large-electrode limit; 
here, however, fringing fields disrupt circuit operation for \(|E_y| >
E_o/2\). In this case, it is possible to apply an input field of 
strength \(|E_y| < E_o/2\), which is adequate to select a bit on the
input+shift-register circuit, but too weak to drive a wrong state on
the shift register. When \(|E_y| > E_o/2\), the field is strong enough to
induce a kink between cells 2 and 4, so that the wrong bit propagates
down the shift register (see insets in the lower left and uppper right of
Figure \ref{fig-basic-clocked-input-ckt}). This result is consistent
with the result previously published for a an unclocked input
QCA circuit \cite{QCAeFieldWriteIn-IEEE}. All calculations use a strong
clocking field (\(E_z = -10 E_o\)), since a clock of this magnitude
results in a strong output bit (\(|P_6| \rightarrow 1\)). 

An even stronger fringing input field (\(|E_y| \gg E_o/2\)), can simply
inject multiple kinks into the shift register, causing all cells to align
with \(E_y\). We do not study the circuit up to this point, since it
is only necessary to find where the \emph{first} failure occurs in order to
determine the limits of operation for this circuit.

\subsection{Clocked Input Circuits - A More Robust Design}
\label{sec:orga8141c5}

\begin{figure}[htbp]
\centering
\includegraphics[width=.9\linewidth]{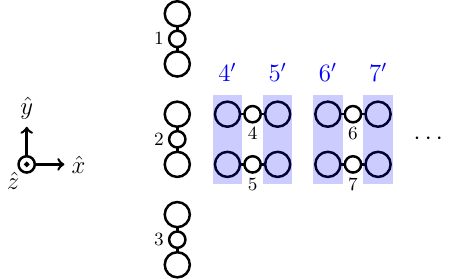}
\caption{An input circuit with a shift register comprised of rotated cells is designed to be immune to fringing fields. Starting with the circuit of \ref{fig-clocked-input-with-limits}, cells 4+ are rotated by 90 and rearranged to function as a shift register. Active dots from cells 4+ are paired (identified with blue rectangles) to form \textit{functional cells} \(4^{\prime}, 5^{\prime}, \ldots,\) etc. \label{fig-input-ckt-immune}}
\end{figure}

The vulnerability of the binary wire cells to the \(y\) component
of fringing input fields suggests a simple modification for more
robustness: the binary wire cells may be rotated by \(90^{\circ}\), as
shown in Figure \ref{fig-input-ckt-immune}. For the rotated cells (4, 5,
6, etc.), \(\hat{y}\perp\vec{a}\) leads to
\(E_{y}\hat{y}\cdot\vec{a}=0\) so that the \(y\)-component of the
fringing input field cannot affect the detuning between their active
states. Thus, the rotated cells provide a binary wire immune to the
\(y\)-component of the field, but the non-rotated cells (1-3) retain
their sensitivity to \(E_y\).

\begin{figure}[htbp]
\centering
\includegraphics[trim=0.0125cm 0 0.09cm 0, clip,width=.9\linewidth]{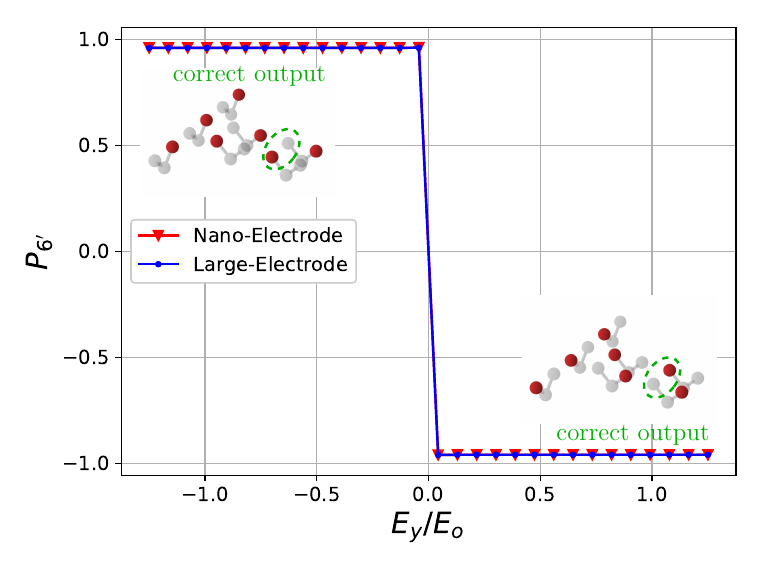}
\caption{The circuit of Figure \ref{fig-input-ckt-immune} demonstrates immunity to fringing fields in the \(y\) direction. \label{fig-immune-ckt-resp-y}}
\end{figure}

Results from the circuit of Figure \ref{fig-input-ckt-immune} are shown
in Figure \ref{fig-immune-ckt-resp-y}. To obtain a result from this circuit
directly comparable to the results of Figure
\ref{fig-basic-clocked-input-ckt}, we group the active dots from the
rotated molecular cells into functional cells, labeled
\(4^{\prime},5^{\prime},\ldots,\) etc. (see Figure
\ref{fig-input-ckt-immune}). In particular, \(P_{6^{\prime}}\), the
polarization of cell \(6^{\prime}\) is plotted, where 
\begin{equation}
P_{6^{\prime}}\equiv \frac{1}{2}\left(P_{6}-P_{7}\right). \label{eq-P6p}
\end{equation}
Notably, the large-electrode \(P_{6^{\prime}}\) response under
unwanted fringing fields in the \(y\) direction is identical to the
nano-electrode \(P_{6^{\prime}}\) response (ideal case), even under
\(| E_y | > E_o/2\). The functional cell polarization \(P_{6^{\prime}}\)
remains unaffected by a strong fringing field with a \(y\)-component,
and the circuit will not fail, even under arbitrarily large
\(|E_y|\) values. As before, the clocking field was set to \(E_z = -10
E_o\) for all calculations.

\begin{figure}[htbp]
\centering
\includegraphics[width=.9\linewidth]{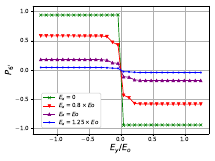}
\caption{A circuit with rotated binary-wire cells demonstrates resilience to the \(x\)-component of fringing input fields. \label{fig-x-fringing-resp}}
\end{figure}

In addition to immunity to fringing fields in the \(y\)-direction, the
input circuit and shift register of Figure \ref{fig-input-ckt-immune}
demonstrates robustness against fringing fields in the
\(x\)-direction. This is shown in Figure \ref{fig-x-fringing-resp},
where \(P_{6^{\prime}}\) is plotted as a function of \(E_{y}\) for
various values of \(E_{x}\). Here, all calculations are in the
large-electrode limit so that \(E_x\) is applied to the entire
circuit. The input field \(E_{y}\) selects the input, and
\(P_{6^{\prime}}\) encodes the correct--albeit weakened--bit until
\(E_x \sim E_o\). Circuit failure requires 
alignment between physical cells 6 and 7. Here, physical cells 6 and 7
align with the field and with one another (i.e., \(P_6 = P_7\)) so
that functional cell \(6^{\prime}\) fails to an undesirable 
state: \(P_{6^{\prime}} \rightarrow 0\). Note that a bit weakened by
an unwanted non-zero \(E_x\) may be strengthened by increasing the
strength of the clocking field, \(|E_z|\); however, it is
simplest and most illustrative to perform all calculations at a single
clocking field strength, chosen here to be \(E_z = -7 E_o\). We use a weaker
clock than before, as it provides a more gradual onset of failure with
increasing \(|E_x|\) than in the case of the stronger \(E_z = -10
E_o\).

\begin{figure}[htbp]
\centering
\includegraphics[trim=0.1cm 0 0.0875cm 0,clip,width=.9\linewidth]{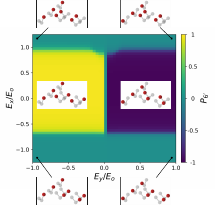}
\caption{A circuit with rotated binary-wire cells demonstrates resilience to the \(x\)-component of fringing input fields. In the yellow and blue regions, the input \(E_y\) selects a bit on the input cells, and the shift register responds properly. In the green regions, a strong unwanted \(E_x\) drives the shift register to failure. \label{fig-x-fringing-resp-pcol}}
\end{figure}

A broader exploration of the effects of the \(x\) component of
the field on the circuit of Figure \ref{fig-input-ckt-immune} is shown
in Figure \ref{fig-x-fringing-resp-pcol}. Here, the blue and yellow
regions are where the shift register functions properly:
\(P_{6^{\prime}}\) is strong and non-zero, despite some unwanted
\(E_{x}\). The green region is where functional cells of the shift
register fail to \(P_{k^{\prime}} \rightarrow 0\) as the physical cells 
simply align with \(E_x\). In the absence of cells 6 and 7, the shift
register (comprised only of cells 4 and 5) would fail when \(E_x \sim
-E_o\): here, the field has to do the work of injecting a kink between
cells 4 and 5 (this is consistent with the result of Figure 8 of
Ref. \cite{QCAeFieldWriteIn-IEEE}). However, with the addition of cells 6
and 7, as in Figure \ref{fig-input-ckt-immune}, the attraction between
neutralizing (mobile) charges in cells 4 and 5 and oppositely-charged
mobile (neutralizing) charges of cells 6 and 7 lowers the energy of
the configuration in which all shift register cells align with the field
component \(E_x\). Thus, the shift register fails for \(E_x \lesssim -0.6
E_o\). Similarly, in the absence of cells 6 and 7, a shift register
comprised only of cells 4 and 5 would be resistant even to some fields
\(E_x > E_o\), since the field must not only inject a kink, but also
overcome the Coulomb repulsion between the mobile charge of cells 1-3
and the mobile charge of cells 4 and 5. Interactions between the (4,5)
cell pair and the (6,7) cell pair lowers the field strength to \(E_x
\gtrsim E_o\) for the onset of shift register failure. Nonetheless, the
circuit tolerates significant non-zero, unwanted \(E_x\) before the
shift register fails: \(- 0.6 E_o \lesssim E_x \lesssim E_o\). Here, all
calculations were performed using a clock of strength \(E_z = -7
E_o\), although a stronger clock may compensate for the effects of \(E_x\)
to some extent.

It is reasonable to design the circuit for immunity to the \(y\)
component of fringing input electrode fields, but only tolerance to
an unwanted non-zero \(x\) component. The \(y\)-component of
fringing fields will be the dominant component applied over the
innterconnecting shift register or even the down-stream logic, since the
applied \(\vec{E}\) will be designed to be dominant in its \(y\)
component.

% ============================================================
% ============================================================
% ============================================================
% SECTION: Discussion
% ============================================================
% ============================================================
% ============================================================

\section{Discussion}
\label{sec:Discussion}

While this paper focuses on clocked molecular QCA inputs, it is worthwhile to contextualize it in the broader QCA paradigm and to briefly discuss open questions and challenges, especially those questions which pertain to this paper. We briefly mention the role QCA could play in general-purpose computing, and then discuss the open challenges of molecular design, synthesis, testing, circuit layout, and bit read-out.

If realized, molecular QCA could provide a room-temperature alternative to CMOS for both logic and memory. Device densities for cells with a \(\sim 1\; \text{nm}^2\) footprint could approach \(10^{14} \; \text{cm}^{-2}\), surpassing present-day CMOS densities by 3-4 orders of magnitude. Bits at this scale are \textit{very} robust: bit energies of \(E_k \sim 420\; \text{meV}\) tower high above the thermal noise floor at room temperature, \(k_B T \sim 26 \; \text{meV}\), eliminating the need for cryogenic temperatures. Molecular QCA operating speeds could approach the THz regime \cite{molecularQCAelectronTransfer}.  Clocked QCA support adiabatic operation for low power dissipation \cite{1999ClockingToth,BennettClocking,LentBlair2011v2}, and this will be important at the aforementioned device densities. Modeling and quantifying power dissipation in molecular QCA candidates is an area of development by other groups \cite{ExponentiallyAdiabaticSwitch2018} as well as our own.

As a general-purpose computing paradigm, molecular QCA could deliver high performance at a low power demand in computing devices from wearables to supercomputers. Molecular QCA circuits could replace CMOS fully, or perhaps provide molecular processors that work along with CMOS, with the advent of suitable write-in and readout techniques for molecular circuits. In QCA, logic and memory may be intermingled, and non-von-Neumann architectures are possible \cite{2003IEEETED-Clocked-mQCA,2016_ICRC,2018_clock_topologies}.

Molecular design, synthesis, and testing are a crucial open questions in molecular QCA. Two-dot, three-dot, and four-dot molecules have been synthesized and probed using scanning-tunneling microscopy (STM), and charge-localized device states have been observed \cite{DFA,2010_STM,STM_ThreeDot_2012}, along with field-driven switching \cite{FieldSwitchDyad2003}. Recently, control was demonstrated over the position of a mobile hole on a double-quantum-dot, bis-ferrocene cation using an atomic force microscopy (AFM) system \cite{AFM_switch_2020}. This molecule is similar to the DFA molecule from Figure \ref{fig-qca-molecules}(a). Additionally, Henderson, et. al. \cite{ZwitterionicNidoCarborane}, synthesized a new molecule in an important class of QCA candidates: zwitterionic, charge neutral, mxed-valence species \cite{Zwitterions2011,2013_zwitterionicQCA_DQD}. This type of molecule is self-doping so that its synthesis avoids the production of counter-ions. This is helpful because counter-ions will localize randomly near the ionic QCA cells, and such stray charge could disrupt circuit operation \cite{2013StrayChargeTougaw}. Progress in molecular QCA would be advanced if stable device charge states and switching/clocking were experimentally demonstrated in a neutral, mixed-valence QCA candidate. The design of neutral, mixed-valence species is an active research area for our research team.
 
To form circuits, non-homogenous patterns of molecules must be arranged on a surface. Promising techniques for molecular QCA circuit layout rely on self-assembly. Some works have explored the concept of bonding molecules to a gold substrate \cite{2013_bisferrocene_wire_layout,2019_Bisferrocene_wire}. Other proposals may combine top-down and bottom-up approaches, such as lithography and self-assembly. DNA rafts could be programmed with specific attachment sites for appropriately-functionalized QCA molecules, allowing for the self-assembled formation of molecular circuits on DNA circuit boards. This could allow suitable control over both the position and orientation of molecules. Top-down techniques such as electron-beam lithography (EBL) can define arrangements for the DNA circuit boards, providing a means for organization at a larger scale \cite{DNARafts2003,DNARaftsExperiment,EBL_DNA_QCA_2005}. Indeed, exquisite control over DNA nanostructures has been demonstrated, both in 2D and 3D \cite{DNA_nanoscale_shapes_and_patterns,DNATiles}. To avoid unwanted effects from stray charge presented by phosphate groups in DNA structures, it may be helpful to develop peptide nucleic acid (PNA) circuit boards \cite{PNA_Nature}. Experimental demonstrations of circuit layout\textemdash including adequate control of both cell placement and rotation\textemdash will be an important step forward in realizing molecular QCA.

QCA circuits may in fact allow for some lack of precision in the positioning and rotation of cells. QCA circuits can tolerate defects (i.e., missing cells, rotated cells, and displaced cells). Circuit robustness can be improved through techniques such as redundancy and clocking \cite{faultToleranceRedundancy,2008_DefectTolerance,2013_bisferrocene_wire_layout}. The circuit model presented here could be used to explore defect tolerance in circuits of limited size, and some approximations could be introduced to explore defects in larger circuits.

Additionally, the read-out of molecular QCA bits is an important technical challenge in QCA. Recent proposals for bit read-out involve the use of single-electron transistors (SETs) or single-electron boxes (SEBs) \cite{SEB_1991,SET_2006,2009_SET_detection,2010_SET}. While exquisitely sensitive, such these devices require low temperatures. A room-temperature read-out technique could be preferable to cryogenic read-out solutions.  Our own research group is investigating room-temperature solutions and may be able to present some novel ideas in the near future.

Not only does this paper anticipate solutions to open questions identified above, but it also is written to \textit{invite} and \textit{encourage} further research and development toward achieving those solutions. The design and synthesis of QCA candidates is of strategic importance, since it is prerequisite to experimental demonstrations of output and input techniques. Circuit layout techniques could be demonstrated with QCA candidates, or even dummy molecules that do not provide QCA device states. Adequate circuit layout techniques will enable straight-forward, robust techniques for bit write-in to molecular QCA circuits. This work also suggests means by which molecular QCA logic could be designed to tolerate strong, unwanted electric fields.

\section{Conclusion}
\label{sec:orgf76010d}
We have extended a model of unclocked input circuits using two-dot QCA
cells to a model of electric-field-clocked, three-dot QCA
molecules. The circuit was treated exactly as a system of \(M\)
three-state QCA molecules. The clocked version of
the originally-proposed input circuit from Ref. \cite{QCAeFieldWriteIn-IEEE}
operates even in the presence of unwatned fringing fields in the
\(y\)-direction. Strong fringing from the input field, however,
can disrupt operation of the shift register portion of the
circuit. Immunity to an arbitrarily large \(E_y\) field component may
be engineered in the shift register by rotating its constituent cells by
90\(^{\circ}\). The rotated cells are no longer sensitive to the
\(y\)-component of the field, which is likely the strongest component
of the field after the clocking component, \(E_z\). Additionally, the
rotated cells demonstrate significant resilience to the \(x\)
component of the field, which is expected to be the weakest component
in the fringing \(\vec{E}\).

This highlights that the use of two orientations for the QCA cells may be useful for making some parts of the QCA circuitry sensitive to an applied input field (\(\vec{a} \parallel \vec{y}\)) and other parts insensitive to the dominant component of the applied input field (\(\vec{a} \perp \vec{y}\)). Importantly, it may be possible to operate clocked molecular interconnects and QCA logic circuits in the presence of unwanted fringing fields from input electrodes.

Other techniques may help to further minimize the effects of fringing input fields on QCA logic. Circuits could be engineered
to maximize the spatial separation between logic and input circuits. Also, logic calculations could occur out of phase with input
operations so that the input field components are minimized during logical computations. Finally, since input arrays are sensitive
to even weak input fields, input fields strengths may be limited so that they cannot disrupt the operation of interconnects and logic.
The extent to which logic tolerates unwanted electric fields is an active area of research for our group.

The robust circuits presented here provide a helpful step
forward in developing viable solutions for bit write-in on molecular
QCA circuits. Techniques for bit write-in are necessary if
high-speed, energy-efficient general-purpose computation is to be
realized using molecular QCA.

\section*{Acknowledgment}

The authors thank Craig Lent of the University
of Notre Dame for discussion on this work.

\bibliographystyle{IEEEtran}
\bibliography{QCAReferences}

\begin{IEEEbiography}[{\includegraphics[width=1in,height=1.25in,clip,keepaspectratio]{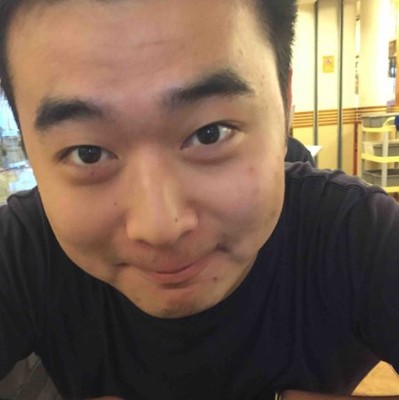}}]{Peizhong Cong} is Ph.D. candidate in the Department of Electrical and Computer Engineering at Baylor University. He received his B.S. (2013) from Tianjin University, M.S. (2016) from University of Rochester. He joined Dr. Enrique P. Blair's research group in 2017. He focus his research on classical molecular computing using quantum-dot cellular automata (QCA); molecular QCA circuit design; and clocked molecular QCA.
\end{IEEEbiography}

\begin{IEEEbiography}[{\includegraphics[width=1in,height=1.25in,clip,keepaspectratio]{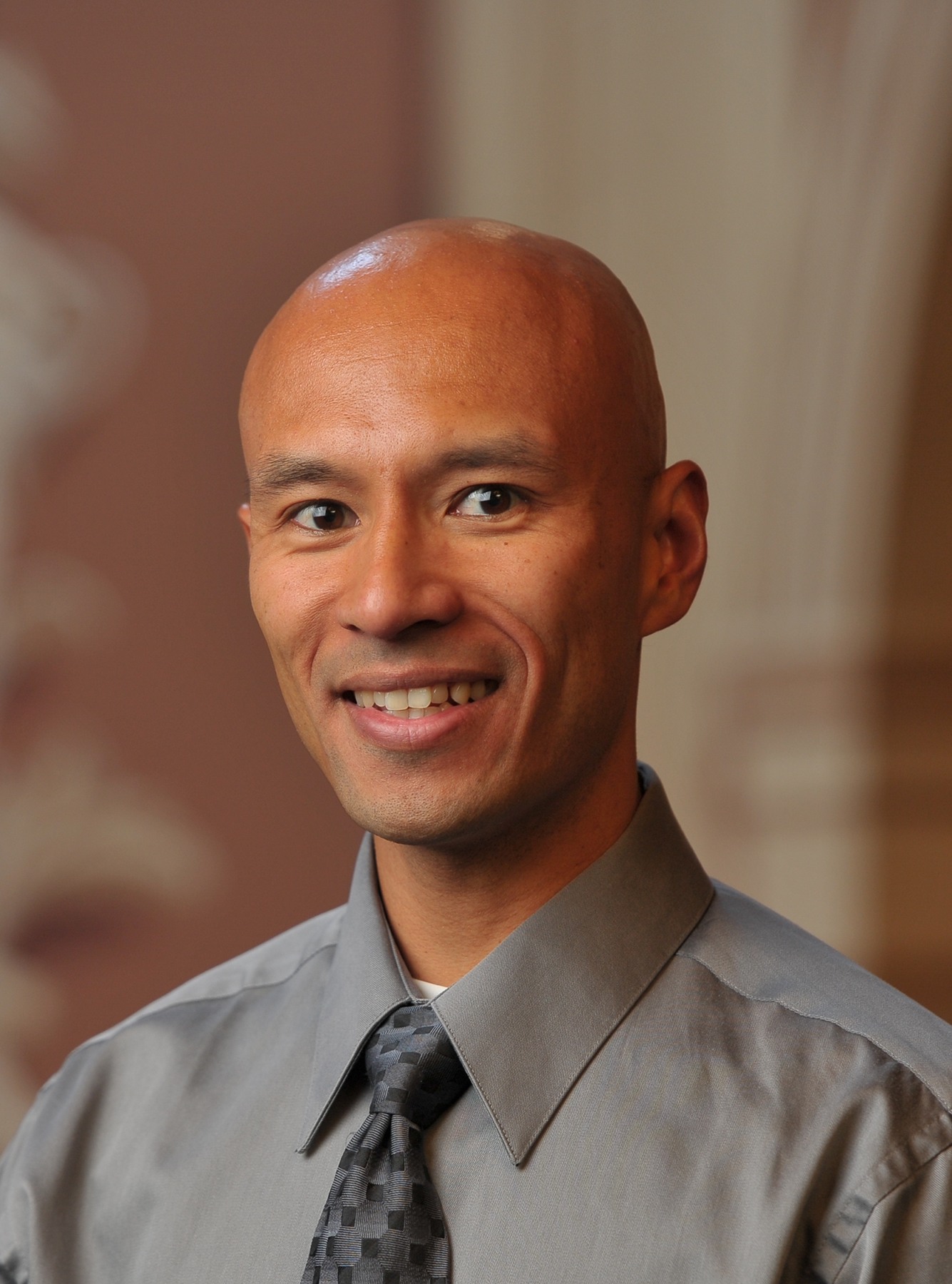}}]{Enrique P.\ Blair (Senior Member, IEEE)} is an associate professor in the Department of Electrical and Computer Engineering at Baylor University. He received his B.S.\ (2002), M.S. (2004),\ and Ph.D.\  (2015) in Electrical Engineering from the University of  Notre Dame. His research interests include open quantum systems and classical molecular computing using quantum-dot cellular automata (QCA); quantum computing using molecular charge qubits; and \textit{ab initio} semiconductor device modeling. He served in the U.S.\ Navy from 2004-2010 as a submarine officer and an instructor at the U.S.\ Naval Academy in the Department of Electrical and Computer Engineering. He has been on faculty at Baylor University since 2015.
\end{IEEEbiography}

\end{document}